# DECIDING HOW TO RESPOND: A DELIBERATIVE FRAMEWORK TO GUIDE POLICYMAKER RESPONSES TO AI SYSTEMS

**Willem Fourie**
School for Data Science and
Computational Thinking,
Stellenbosch University
*willemf@sun.ac.za*

## Abstract

The discourse on responsible artificial intelligence (AI) regulation is understandably dominated by risk-focused assessments and analyses. This approach reflects the fundamental uncertainty policymakers face when determining appropriate responses to current, emerging and novel AI systems. In this article, we argue that by operationalising the concept of freedom – the philosophical counterpart to responsibility – a complementary approach centred on the potential societal benefits of AI systems can be developed. The result is a discursive framework grounded in freedom as capability and freedom as opportunity, which represent the two main intellectual traditions of interpreting freedom. We contend that the complexity, ambiguity and contestation involved in regulating AI systems make a deliberative paradigm more useful than the conventional technical one. The resulting framework is structured around coordinative, communicative and decision spaces, each with sequential focal points and associated outputs.

## 1. Introduction: The challenge of regulating AI systems

The regulation of artificial intelligence (AI) systems is exceptionally challenging (Mennella et al., 2024; Shetty et al., 2025; Sousa Antunes et al., 2024; Walter, 2024; Nordström, 2022).

This is in part because the concept of AI systems covers a broad range of system components and types. Furthermore, these systems are deployed in many organisations and sectors, and their use spans multiple linguistic, cultural and geographical boundaries. These systems also tend to evolve fast, making it close to impossible to develop and institutionalise a lasting regulatory regime. It has been argued that uncertainty regarding definitions, outcomes of AI implementations and a disconnect between the pace of AI and policymaking lie at the core of the challenge (Nordström, 2022).

The extent and nature of the impact of AI systems, particularly those based on general-purpose AI models, are difficult to predict (Diab, 2024). The overuse of such systems illustrates this difficulty. While linked to improved productivity, it is possible that, over the medium to long term, increased use of AI technologies could be linked to deskilling (Watermeyer et al., 2024; Woodruff et al., 2024).

The challenge of regulating AI systems effectively and dealing with the ethical issues associated with their development and deployment (Birhane, 2021; Floridi and Cowls, 2019; Mittelstadt et al., 2016) is particularly pronounced in the case of frontier AI systems. These systems are defined as 'highly capable foundation models that could have dangerous capabilities sufficient to pose severe risks to public safety and global security' (Anderljung et al., 2023). The rapid proliferation and evolving capabilities of frontier AI systems pose significant challenges to policymakers (Anderljung et al., 2023), particularly in contexts where malicious actors use them to cause large-scale harm (Shavit et al., 2023).

Conventional approaches to regulating AI can be clustered together under the heading of 'responsible AI' (Goellner et al., 2024; Batool et al., 2023; Schiff et al., 2020; Bach et al., 2025). These approaches operationalise the philosophical concept of responsibility and focus on identifying and mitigating risk. This orientation is also evident in the EU's AI Act, in many respects the global signature piece of legislation when it comes to regulating AI.

At the core of the issue for policymakers is that they face uncertainty when deciding whether and how to act, and if they opt to act, they are confronted with a decision on which policy instruments to use. According to Hood's classification (1983), policymakers have policy options that cover knowledge and communication (nodality), formal rules (authority), economic incentives (treasure) and their own institutions and personnel (organisation). Within each of these components, various options are available. The authority component, for example, includes instruments with various levels of authority, ranging from laws and regulations to softer instruments such as strategies and action plans. Vedung, Bemelmans-Videc & Rist (1998) distinguish between 'carrots, sticks and sermons'. The sticks component largely overlaps with Hood's authority component, with carrots overlapping with treasure and partly with organisation, and sermons mostly with nodality.

To aid policymakers in deciding how to respond, we developed a deliberative framework centred on the societal impact of AI systems. This framework complements existing approaches in two respects. Firstly, the framework shifts the philosophical focus from responsibility, and thus risk mitigation, to freedom, and thereby to the enablement of human capabilities and opportunities. Put differently, this framework uses societal benefit rather than risk as orientation. Secondly, the framework shifts the focus from a technical rationality of policymaking to a deliberative rationality. Policymaking as deliberative project is particularly suited for complex, ambiguous and contested policy problems, such as those posed by AI systems (Sanderson, 2002; Peters, 2004; Fischer and Gottweis, 2013). The resultant deliberative framework is aimed at providing policymakers with a structured approach to understanding the societal impact of AI systems, and using this information as a basis for deciding whether and how to respond with which policy instrument.

The framework is grounded in the discursive policymaking paradigm, and builds on a growing body of knowledge (Nordström, 2022) which treats policymaking as a process of structured argumentation and meaning-making rather than a linear technocratic project (Fischer and Forester, 1993; Hajer, 2003; Schmidt, 2008, 2010; Durnová et al., 2016). It draws on the distinction in discursive institutionalism between a coordinative space (where actors set rules and procedures) and a communicative space (where meanings are contested and co-defined), adding a decision space (where closure and accountability occur) (Schmidt, 2010).

The framework operationalises the concept of freedom in a way that broadens its use in the current discourse on AI regulation. Consider, for example, OpenAI's *AI in America: OpenAI's Economic Blueprint* (OpenAI, 2025), released in January 2025. This document outlines the company's proposals for regulating AI in the United States. The concept of freedom explicitly serves as the conceptual foundation of the document. OpenAI argues that 'straightforward, predictable rules' will ensure 'greater freedom for everyone', emphasising the 'individual freedoms at the heart of the American innovation system'. Central to this vision is the 'freedom for developers and users to work with and direct [OpenAI's] tools as they see fit', opposing the use of AI tools by governments to 'amass power and control their citizens'. OpenAI positions freedom as a democratic value integral to the United States' global economic leadership. Users, developers and governments are expected to exercise the 'freedom to responsibly direct and work with AI tools', creating a self-reinforcing cycle of innovation and, ultimately, prosperity. While OpenAI's use of freedom largely focuses on so-called negative freedom, as will be defined in more detail below, our framework integrates both positive and negative traditions of understanding freedom.

The proposed framework responds in many respects to the recent analysis by Sioumalas-Christodoulou and Tympas (2025). In this analysis, the topics covered in 43 national AI strategies are compared with the topics in global AI indices and metrics. The authors find a 'critical misalignment' between these two datasets, specifically in relation to the underrepresentation of topics related to social impact in global AI indices and metrics. The danger, according to the authors, is that technological innovation might ultimately be prioritised over 'societal and ethical concerns'. While national AI strategies do reflect on the societal requirements and impacts of AI systems, and, by extension, on their ethical implications, these topics are largely absent in indices that measure various topics related to the development of AI systems. The proposed framework is particularly aligned with the authors' contention that the 'direct relevance of these [societal and ethical] challenges to the Sustainable Development Goals' requires indices that can 'specifically measure AI's impact in these areas'.

The framework is also aligned with a recent analysis of just over 100 AI safety evaluations (Griffin and Jacobs, 2025). Among their preliminary findings is the fact that 'no single AI evaluation, or AI meta-evaluation' can reliably report on the societal impact of an AI system. They cite the role of AI systems in misinformation campaigns as an example. AI-assisted misinformation is not merely about 'evaluating the average factuality of AI models' outputs or even tracking the effects of AI use on individual users'. Rather, the actual impact also includes broader societal trends, such as changes in people's opinions and the socio-political impact thereof.

The rest of the article is divided into three parts. The first part explains the shift from responsibility as philosophical orientation to freedom. The second explains the shift from policymaking as technical to deliberative project. The third and final substantive section describes the resultant components of the proposed deliberative framework to guide policymaker responses to AI systems.

## 2. From responsibility to freedom

The framework complements existing risk-based approaches to regulating AI, which in our view operationalises the philosophical concept of responsibility, by operationalising the concept of freedom. The result is a framework that uses the expansion of human capabilities and opportunities as orientation. To understand this conceptual move, we start by discussing responsibility and risk as basis for much of the current discussions on regulating AI systems.

### 2.1. Responsibility and risk

The concept of 'responsible AI' is of central importance in the discourse on the regulation of AI systems and has currency, albeit in some cases indirectly, in practical debates and approaches to regulating AI systems. The responsible AI discourse understandably focuses on mitigating the risks posed by AI and does so through various principles, guidelines and regulatory proposals. The result, in our reading, is an approach to regulation that centres on AI systems rather than human or societal well-being, even though the ultimate goal is indeed to mitigate and address risks in service of societal well-being.

When referring to approaches that centre on AI systems, what is meant is policies and frameworks that use AI systems, their capabilities and risks as the explicit frame of reference, even though the ultimate goal might be to maximise human wellbeing. This is different from using human needs and capabilities, at the explicit level, as framing mechanisms. This section outlines one reading of the contours of the responsible AI discourse and practice, highlighting what seems to be its risk-centred orientation.

At the definitional level, Goellner et al. (2024) observe in their review of 254 research papers that responsible AI is most frequently framed using system-centric terms such as trustworthy, ethical, explainable, privacy-preserving and secure AI. While the terms identified in the study aim to ensure safety, fairness and transparency, they also reveal an implicit assumption: the focus is often on making AI systems predictable and controllable rather than explicitly focusing on the connection between AI systems and individual and societal benefits. For example, explainable AI emphasises understanding decisions made by systems, often as a countermeasure to the opacity of algorithms, whereas trustworthy AI focuses on preventing malfunctions rather than fostering trust in AI's contributions to human development. In the context of the rapid advancement of sophisticated AI systems it is noteworthy that the study does not provide detailed reflection on some of the technical challenges related to explainability.

This orientation around the risk inherent to an AI system can also be seen in AI governance principles. An analysis of peer-reviewed research focusing on these principles across sectors found that responsible AI is consistently associated with addressing risks related to transparency, privacy, accountability, bias and security (Anagnostou et al., 2022). The centrality of risk associated with AI systems is perhaps most pronounced when examining responsible AI governance. In their review, Batool et al. (2023) demonstrate that principles of transparency, bias mitigation, accountability and security dominate the responsible AI governance literature, with limited attention given to the systemic potential of AI to enhance human and societal well-being. Most governance approaches focus on mitigating the potential harms of AI applications in high-risk sectors such as healthcare or finance, often overlooking the broader potential of AI to transform and elevate human experiences across less regulated domains. Schiff et al. (2020) similarly highlight how AI governance frameworks tend to emphasise the prevention of data breaches, algorithmic biases or other system failures, sidelining considerations of how AI could create conditions for societal flourishing.

Also when implementation is considered, the emphasis remains on risk mitigation, with human well-being often cited as a motivation yet rarely forming the central analytic lens. Bach et al. (2025) found that nearly 80% of empirical studies on responsible AI applications focus on deployment in high-risk contexts. This operational focus narrows the scope for exploring AI's potential to positively impact society, reducing the role of governance primarily to minimising damage rather than enabling more significant positive social outcomes.

In practice, the most prominent policy instrument aimed at regulating AI is undoubtedly the European Union's (EU) Artificial Intelligence Act (Chun et al., 2024; Roberts et al., 2023). It is influential within the EU and is likely to shape global policy, much like the EU's General Data Protection Regulation (Tarafder and Vadlamani, 2025). The Act famously adopts a risk-based classification model, categorising AI systems into four groups: unacceptable, high, limited and minimal risk, each with distinct compliance obligations.

Although the risk-based approach should be affirmed and even deepened (Ebers, 2024), the EU AI Act lacks a mechanism for assessing societal benefits alongside risks. For examples of what a more prospective or enabling approach looks like in practice it is instructive to consider other pieces of adopted or proposed EU legislation, such as the European Innovation Act, the European Research Area Act and perhaps even the European Chips Act. From this perspective it is possible to argue that the EU AI Act focuses heavily on risk avoidance without offering sufficient tools for evaluating or integrating the potential positive impacts of AI systems. Others (Novelli et al., 2024) further critique the Act's emphasis on fields of application over specific implementation scenarios, which limits the law's responsiveness to context-specific effects. The risk bias in the EU AI Act aligns with other analyses of AI strategies, particularly in relation to the extent to which societal impact or 'social good' is reflected substantively in these strategies. An analysis of national AI strategies in the EU, for example, found limited reflection on 'the use of AI to tackle social problems' (Foffano et al., 2023). While topics related to AI's impact on society are not absent in national strategies, and indeed more prevalent than in global AI indices and metrics (Sioumalas-Christodoulou and Tympas, 2025), they remain framed within a system- and risk-centric approach to regulation.

Importantly, even the most forward-looking features of the Act, such as AI regulatory sandboxes, which are rightly acknowledged as a policy approach to be emulated (Boura, 2024; Buocz et al., 2023), are framed as mechanisms for managing uncertainty and reducing compliance burdens, not as tools for proactively maximising societal benefit.

The impact assessment conducted by the European Commission (2021) in preparation for the legislation reinforces this reading of the EU Act. The assessment frames the regulation as a response to six policy problems, of which two (1 and 2) focus explicitly on the risks posed by AI, one (3) on challenges related to ensuring compliance, one (4) on regulatory uncertainty and one (6) on legal fragmentation. Only one policy problem (5) points to the risks of not making the most of AI for societal benefit, and even this is framed narrowly in terms of the EU's global competitiveness.

### 2.2. Freedom and societal impact

By identifying a risk bias in the current responsible AI discourse, we are not arguing against the value of this approach. Rather, the argument presented in this article is that the risk-based approach can benefit from additional orientations. The proposed framework suggests a related yet alternative point of departure, namely the concept of freedom.

As discussed in the introduction, freedom is already used as a point of reference for regulating AI systems. Yet, as we will show in this section, it is a much more comprehensive concept than suggested by key industry players. In this respect, the framework challenges arguments that pure self-regulation will be sufficient to ensure the maximal aggregate societal impact of AI systems. In fact, the move towards self-regulation holds the danger of 'regulatory gifting' (Papyshev and Yarime, 2024), where policymakers' well-intentioned moves 'reduce or reorient regulators' functions to the advantage of the regulated and in line with market objectives on a potentially macro scale'. The danger of 'regulatory gifting' looms particularly large in settings where 'a non-binding and unenforceable principles-based approach to the regulation of AI' is followed.

Before proceeding to our discussion of the concept of freedom, we first discuss the close relationship between freedom and responsibility. The philosopher Kant's definition of enlightenment provides the starting point for understanding this relationship. According to Kant, enlightenment should be understood as the individual's liberation from their self-imposed immaturity, where immaturity is understood as the inability to use one's own reason without guidance from others (Kant, 1784). This immaturity is self-imposed because it stems not from a lack of rational capacity but from a failure to make the courageous decision to use one's rational capabilities.

Kant's understanding of the connection between freedom and responsibility is made explicit in his first formulation of the categorical imperative – the most influential construct in his ethics. According to this formulation, one should only act 'according to that maxim whereby you can at the same time will that it should become a universal law' (Kant, 1785). Freedom is not a licence to do as one likes. Rather, freedom is intimately linked to the well-being and conduct of others and, ultimately, to what constitutes a good society. The enlightened person, therefore, is the person who uses their rational capabilities to identify and apply principles that serve both personal and collective well-being.

In the second formulation of the categorical imperative, Kant makes clear that the close relationship between freedom and responsibility is anchored in the dignity of all human beings. In this formulation, one should 'act that you use humanity, whether in your own person or in the person of any other, always at the same time as an end, never merely as a means' (Kant, 1785). The maxims on which one's actions are based should be generalisable exactly because human beings are ends in themselves. Kant therefore also describes a person as something 'beyond a price', as no individual can be replaced by anything else. Humans are entitled to freedom but must, at the same time, exercise it in ways that respect the dignity and freedom of others. From this perspective, responsibility is the philosophical counterpart to freedom.

While Kant's work provides a foundation for understanding the interplay between freedom and responsibility, subsequent theorists have expanded and clarified the concept of freedom. Among these, Berlin's distinction between two concepts of freedom, namely negative and positive freedom, remains foundational. In his inaugural lecture *Two Concepts of Liberty* (Berlin, 1958), Berlin identifies negative freedom as the absence of external interference in one's choices and actions, or 'the degree to which no man or body of men interferes with my activity'. Rooted in classical liberalism and articulated by thinkers such as Locke and Mill, negative freedom ensures that individuals can pursue their values without coercion. Concepts such as individuality, rationality, dialogue and rights are central to this intellectual tradition.

Positive freedom chiefly concerns the capacity for self-mastery. This requires overcoming internal obstacles such as ignorance, irrational desires or psychological compulsions to achieve personal autonomy. Berlin associates positive freedom with philosophical traditions rooted in Rousseau, Kant and Hegel. This concept has also been interpreted as emphasising that individual freedom depends on conditions beyond individual control. Concepts such as sociality, justice, solidarity and equality are essential to understanding this form of freedom.

Building on Berlin's distinction, Sen reinterprets positive freedom in terms of individual capabilities. In this way, he connects freedom with developmental economics and social choice theory. For Sen, negative and positive freedom are not mutually exclusive but interdependent, as realising substantive freedom requires both the absence of interference and the presence of enabling conditions (Sen, 2002).

Other scholars have further explored this interplay, including theories of communicative freedom. The concept of communicative freedom has its origins in the field of social ethics, with the theologian Huber expanding on the concept initially proposed by the philosopher Michael Theunissen in the late 1970s (Huber, 1985). Huber argues that individual and social dimensions of freedom are mutually dependent and, in fact, have the same source. While his proposal is fundamentally theological (Fourie, 2012), he does echo Kant's contention that human dignity is to be viewed as at the basis of the connection between different conceptions of freedom, and as an end in itself rather than a means to an end. By placing the worth, dignity and freedom of the individual outside the realm of control of any societal actor, Huber lays the foundation for a concept of freedom that views its individual and social dimensions as complementary and not contradictory. This, in turn, allows for a conception of freedom that is not bound to either the liberal or social traditions of interpretation.

When considering operationalising freedom to improve AI regulation, Taylor offers a particularly useful reformulation of Berlin's ideas, echoing also elements found in the thought of Huber. Taylor reframes freedom in terms of freedom as opportunity and freedom as exercise (Taylor, 1985). Freedom as opportunity refers to the options available to a person – what they can do, regardless of whether they act on these options. Freedom as exercise, on the other hand, concerns the ability to actualise chosen options and requires self-determination. Taylor's distinction shows the absence of constraints without substantive opportunities is of limited value, just as opportunities are irrelevant without the capacity to act on them.

Synthesising these perspectives, the concept of freedom can be operationalised for the proposed framework as freedom as capability and freedom as opportunity. *Freedom as capability* focuses on the capabilities of individuals that enable them to live dignified lives. Freedom as capability covers how AI systems support and expand capabilities internal to an individual. *Freedom as opportunity* focuses on the absence of external constraints that limit action as well as on the actual opportunities available. This dimension evaluates whether AI systems reduce or increase the options available to individuals. Whereas capabilities address internal conditions, opportunities address the external environment surrounding individuals.

## 3. From technical policymaking to deliberative policymaking

The proposed framework complements existing approaches to regulating AI systems in two respects. The first, as discussed in the previous section, is a focus on freedom as point of orientation. The second, as will be discussed in this section, is by adopting a discursive, rather than purely technical, approach to policymaking. As we will argue in this section, the complexity, ambiguity and contestation associated with deciding how to regulate AI systems makes this approach particularly useful.

### 3.1. Policymaking as technical project

Two rationalities continue to dominate policymaking research and practice: technical and deliberative rationalities. The genesis of technical rationality, also called instrumental or positivist rationality, is often traced back to Harold Lasswell's *The Decision Process: Seven Categories of Functional Analysis* (Lasswell, 1956). Through functions such as intelligence, recommendation, prescription, invocation, application, appraisal and termination, Lasswell conceives policymaking as a linear and largely functional process. Around the same time, other technical approaches converged with Lasswell's thinking. Herbert Simon, for example, famously introduced the idea of bounded rationality. While critiquing overly positivist views of human rationality, Simon still frames policymaking as a systematic and ultimately technical decision process (Simon, 1957).

In the 1960s, theorists such as Downs (1967) and Dror (1967) turned the lens towards bureaucrats or policymakers themselves. Even Dror's definition of a bureau, a type of organisation, demonstrates technical rationality at play. A bureau, for example, should be large enough that its highest-ranking members know less than half of all members personally. His central hypothesis is framed in technical terms and concerns how officials attain their goals. Taken together, officials are 'utility maximisers' who seek to attain their goals rationally, given their limited capacities and the cost of information. They pursue a complex set of goals, which includes power, income, prestige, security, convenience, loyalty, pride and the desire to serve the public.

Conceiving policymaking as a process based on technical rationality has led to many innovative and, in some cases, groundbreaking approaches. In his influential book *The Logic of Collective Action*, Olson (1965), for example, coins the free rider problem, showing how individuals will not contribute to collective goods without appropriate incentives. He frames the problem of collective action failures as a design issue to be solved technically through incentives. Niskanen, during the same period, conceptualised policymakers as budget maximisers, thus opening the way for quantifying the effects of policymakers' actions (Niskanen, 1971).

The New Public Management movement can be seen as a later expression of technical rationality. This is particularly clear in the extent to which values in public administration are instrumentalised. According to Hood (1991), for example, New Public Management should be based on three clusters of values, with frugality, rectitude and resilience defining the standards of success in policymaking. These ideas were also articulated by Aucoin (1990) and Pollitt (1990).

Elements of the evidence-informed policymaking movement continue to produce innovations and limitations informed by a technical rationality. The standard for determining the impact of policies impact, and thus improving policymaking, is the randomised control trial (RCT) (Kremer, 2003; Banerjee and Duflo, 2009; Banerjee, 2015). RCTs typically randomly assign individuals to control and treatment groups, with the control group not exposed to the policy intervention. The ultimate goal is to identify causal effects to generate generalisable evidence.

However, technical rationality in policymaking encounters limits when policy problems are ambiguous or contested, or when policy impacts are difficult to determine in the short to medium term (Sanderson, 2002). This line of criticism has been present from the onset, notably in Lindblom's theory of incrementalism, expressed in his famous 'muddling through' approach (Lindblom, 1959; Lindblom, 1979). In our view, this is certainly the case for interventions involving powerful, and at times general-purpose, technologies such as artificial intelligence.

### 3.2. Policymaking as deliberative project

Challenges to the technical view are often framed through the 'argumentative turn' introduced by Fischer and Forester (1993). This turn, based on the philosophy of Habermas (1990) and Foucault, places discourse at the centre of the policymaking process. It is not merely about acknowledging the importance of discourse but involves what some term a post-positivist approach (Peters, 2004), where definitions and understandings of impact are co-constructed in a deliberative process. Discourse thus becomes ‘the site of production of knowledge and power’ (Durnová et al., 2016). When conceiving policymaking as a deliberative project, it can be understood as ‘an ensemble of ideas, concepts and categories through which meaning is given to social and physical phenomena, and which is produced and reproduced through an identifiable set of practices’ (Hajer, 1995).

The argumentative turn and its associated deliberative rationality have been shown to be particularly useful for complex and riskier policy problems under conditions of uncertainty (Ney, 2009). According to Fischer and Gottweis (2013), overly technical approaches to such ‘messy’ problems ‘have in fact frequently compounded the task, as they have themselves become a source of ambiguity and thus uncertainty’. The result is that policy analysis, and policymaking more generally, cannot rely on deductive reasoning alone, as such reasoning fails to address the nonlinear nature of messy policy problems.

Schmidt, in her work on discursive institutionalism, discusses two types of discourse, which we will expand on in subsequent sections (Schmidt, 2008). Coordinative discourse involves the actors who will ultimately make decisions and covers processes related to ‘policy construction’. In the discursive framework discussed later, this is operationalised as two phases: the first, during which the rules of the discourse are set, and the third, during which final decisions are made. Schmidt also discusses communicative discourse, which ‘consists of the individuals and groups involved in the presentation, deliberation and legitimation of political ideas to the general public’. In our deliberative framework, this is the centrepiece and constitutes the second of three phases. During this phase, definitions and conceptions of impact are co-constructed, though this will often involve deep and enduring disagreements.

Central to policymaking as a discursive project is the notion of participation. At the surface level, this involves the breadth of participation and related power structures. For example, who decides which actors are entitled to participate? At a deeper level, as discussed by Fischer (2006), participation can be measured through its effects. Participation enables instrumental effects when it helps achieve policymaking goals that individuals could not reach on their own. Developmental effects arise when participation improves citizens’ capacities. Intrinsic effects, particularly relevant in participative democracies, refer to the inherent value of meaningful participation. The focus of our discursive framework is on participation that enables both instrumental and developmental effects.

## 4. A deliberative framework to guide policymaker responses to AI systems

Based on freedom as philosophical framing and deliberative policymaking as paradigm, we will discuss the components of the proposed framework in this section.

The framework is intended to guide policymakers in deciding how to respond to existing, emerging and novel AI systems operating in their jurisdictions. It is intended to complement, not replace, existing responsibility- and risk-based approaches, such as the EU AI Act, by providing a structured method for considering potential societal benefits (Chun et al., 2024; Ebers, 2024) under conditions of high uncertainty.

In our view, the framework could be useful in several policymaking contexts and processes. In early-stage regulatory development, it can help systematically map potential societal impacts beyond immediate risks. During ex-ante impact assessments for specific regulations or deployments, it can provide a guide for evaluating potential positive and negative effects in multiple domains and in this way it can be used as basis to determine further policy responses. The framework could also improve AI safety evaluations, specifically addressing known weaknesses in assessing broader societal consequences (Griffin and Jacobs, 2025).

Despite multiple potential applications, the framework is not a benchmarking or certification tool. Rather, its value lies in providing a structured, transparent and replicable method for deliberation on the societal impact of AI systems. The aim is to reach a negotiated, if contested, understanding of societal impact, to be read alongside risk-focused assessments. In this way the framework resonates with the ordoliberal approach to AI governance (Hälterlein, 2025), where a structured ‘third way’ between regulatory overreach and laissez-faire regulation is sought.

As will become clear in the next sections, completing the framework will almost certainly be a time-consuming exercise, though in our view not prohibitively time-consuming. Yet is also our view that the time devoted to applying this framework should over the medium benefit policymakers and ultimately society by enabling better application of policy instruments. Figure 1 provides a summary of the main components of the framework.

Table 1: Summary of the deliberative framework to guide policymaker responses to AI systems

| Component | Focus | Sequential Steps | Outputs |
|---|---|---|---|
| **Coordinative space** | Establishes procedural legitimacy | - Define mandate and limits<br>- Select participants<br>- Set rules of engagement<br>- Appoint facilitators | - Terms of reference<br>- Stakeholder map<br>- Process charter<br>- Facilitator mandate note |
| **Communicative space** | Substantive deliberation on societal impact | - Orient participants<br>- Define process<br>- Define principles<br>- Make qualitative judgements | - Briefing dossier<br>- Deliberation roadmap<br>- Working definition matrix<br>- Argument log |
| **Decision space** | Ensures closure and accountability | - Review fairness and representation<br>- Synthesise stakeholder judgements<br>- Closure via agreed rules<br>- Report decisions and disagreements | - Procedural compliance note<br>- Synthesis map<br>- Decision record<br>- Public deliberation report |

### 4.1. Orientation

The framework is oriented philosophically by the concept of freedom, which has been operationalised as freedom as capability and freedom as opportunity. By doing so, the framework takes societal benefit as its point of orientation. Freedom as capability focuses on the conditions internal to an individual, organisation or institution that enable them to act in the world. Freedom as opportunity focuses on the opportunities in the external environment that enable individuals to actualise their capabilities.

To translate this philosophical concept into a practical deliberative tool, the framework draws on the Sustainable Development Goals (SDGs) as globally recognised goals of societal development. While acknowledging that the SDGs were not formulated with AI governance specifically in mind, and recognising ongoing debates about their implementation and measurement, they currently represent the most widely accepted international articulation of societal development objectives. Their applicability across diverse national contexts and widespread use in domains like corporate sustainability reporting (Nicolò et al., 2023) make them a practical and legitimate starting point for defining impact.

The SDGs are typically organised around the so-called ‘five Ps’: people (SDGs 1-5), planet (SDGs 6, 12-15), prosperity (SDGs 7-11), peace (SDG 16) and partnership (SDG 17) (Tremblay et al., 2020). In applying this structure, the framework focuses on four of the five pillars, deliberately setting aside ‘partnership’ as it primarily concerns mechanisms of international cooperation rather than societal impact.

When considering capabilities the SDGs provide at least the following deliberative starting points: poverty reduction (SDG 1), access to food (SDG 2), health (SDG 3), education (SDG 4), water (SDG 6), energy (SDG 7), and adequate housing within sustainable settlements (SDG 11). For opportunities the SDGs provide the following starting points: economic participation and decent work (SDG 8), infrastructure and innovation (SDG 9), reducing inequalities (SDGs 10 and 5), environmental sustainability (SDGs 13, 14, 15), and peaceful, just and effective institutions (SDG 16), which includes aspects of safety and security. SDG 17 (‘partnership’) is excluded as its targets primarily concern international cooperation mechanisms (e.g., trade, aid, finance, technology transfer). While relevant for global development, these topics fall more within the domain of international relations than societal impact.

It is important to distinguish this framework’s use of the SDGs from the growing body of research on how AI contributes to achieving the SDGs directly. A highly cited piece of research in this context is Vinuesa et al. (2020), who have shown, for example, how AI systems can accelerate the achievement of 134 of the 169 targets of the SDGs while hindering the achievement of 59 targets. In the framework, the SDGs fulfil a related yet different function. In the context of weakening consensus on global development goals and global cooperation, the SDGs remain one of the few lists of goals with broad acceptance across national borders. Viewed in this way, the SDGs provide a useful starting point for applying the two dimensions of freedom used in this framework

### 4.2. Paradigm

The framework uses deliberative policymaking as paradigm. Whereas technical approaches to policymaking are particularly well-suited for problems with clear definitions and linear responses, deliberative approaches emphasise inclusive dialogue under conditions of uncertainty. The framework is built on the assumption that policy responses to AI systems require a different paradigm. High levels of uncertainty regarding AI's societal impact, and ambiguity about the exact policy problem to be addressed, necessitate a different paradigm.

The application of this paradigm requires substantive participation of a wide range of stakeholders, at least enabling instrumental and developmental effects (Fischer, 2006). At least the following groups need to be represented when making use of this framework:

- *Policymakers*: These individuals have a key role in defining the coordinative space and will ultimately be tasked with coming to a final determination in the decision space.
- *Affected parties*: Affected parties should be self-identified in response to an open invitation.
- *Domain experts*: Individuals with specialised academic or technical knowledge relevant to the technologies being assessed will need to participate. These individuals should not be employed by the entities responsible for developing the AI system under evaluation.
- *System developers*: This group includes individuals or teams involved in the design, development, deployment or operation of the AI system under evaluation. They should possess insight into the capabilities, intended use, users and operational parameters of the system under evaluation.
- *Interpreters*: To make substantive participation possible, interpreters with the ability to clarify particularly technical aspects of the system under review as well as the ability to articulate the input of affected parties will be required.

### 4.3. Components

The framework consists of three spaces which are activated sequentially. Participation in the communicative space is the broadest. The coordinative and decision spaces are to be regarded as connected, in the sense that the coordinative space needs to enable a definition of the 'rules of the game' that make it possible to conclude the process in the decision space.

#### 4.3.1 Coordinative space

The focus of the coordinative space is procedural and not substantive. This space establishes the legitimacy of the process aimed at determining the societal impact of an AI system and also sets the boundaries and rules of engagement. This corresponds to the coordinative space in discursive institutionalism (Schmidt, 2008, 2010) where rules, boundaries and participant selection are established before substantive deliberation begins (cf. also Hajer, 2003).

The coordinative space has the following sequential focus areas:

- *Mandate and limits*: Define legal frameworks, scope and deadlines of the process.
- *Participant selection*: Determine stakeholder groupings and representation criteria.
- *Rules of engagement*: Set facilitation rules, evidence standards and fallback procedures when consensus cannot be reached.
- *Appointment of facilitators*: Appoint legitimate facilitators to guide the communicative space.

While policymakers make up the core participants of the coordinative space, participants should also include representatives from the affected parties, domain experts and system developer groupings.

#### 4.3.2 Communicative space

The focus of the communicative space is substantive deliberation.

This space has the following sequential focus areas:

- *Orientations*: Participants are oriented in four respects. Firstly, they are oriented procedurally regarding the mandate and limits of the process and the rules of engagement. Secondly, they are oriented technically with regard to the AI system under review. Thirdly, participants are oriented normatively with regard to freedom as capability and freedom as opportunity serving as the proxy definitions for societal impact. Fourthly, participants are oriented substantively with regard to the SDGs as deliberative starting points within each dimension of freedom.
- *Process and principles*: Participants co-define the process steps to be followed in the subsequent focus areas.

- *Definitions*: Participants deliberate on how freedom as capability and opportunity should be defined in the context of the AI system under review. In each case, the identified SDGs are used as deliberative starting points.
- *Judgements*: For each substantive component (e.g. education, health, work, environment) under each normative dimension (capability, opportunity), participants provide qualitative judgements (e.g. clearly constrains – mixed – clearly enables). The aim is not quantification but the production of an argument log and, where possible, consensus.

All relevant stakeholders should participate substantively in this space.

4.3.3 Decision space

The focus of the decision space is closure and accountability. Following the deliberation of the communicative space, the process moves into the decision space, where closure and accountability are ensured.

This space has the following sequential focus areas:

- *Review*: Facilitators review the process followed during the communicative space and provide their judgement on whether the rules of engagement defined in the communicative space have been followed. Interpreters review the judgements produced and assess whether stakeholder input was fairly represented in discussions. At this point decisionmakers may opt to re-constitute the communicative space.
- *Synthesis*: Decisionmakers apply the rules set in the coordinative space to synthesise judgements produced in the communicative space. Where needed, they may request further input from selected stakeholders to clarify judgements.
- *Closure*: Decisionmakers apply the pre-agreed decision rules (e.g. consensus, majority, or fallback) while ensuring minority views remain on record to decide on a course of action.
- *Reporting*: Decisionmakers publish a report of the final decision, with justificatory reasons and unresolved disagreements.

The decision space should have the same participants as the coordinative space.

**4.4. Output**

Operationalising the components of the framework, the following outputs are identified for each of the spaces and their focus areas. In addition to documenting the process, these outputs provide transparency and legitimacy to the process.

4.4.1 Coordinative space

- *Mandate and limits*: A terms of reference document outlining scope, timeframe, authority and boundaries of deliberation.
- *Participant selection*: A stakeholder map and inclusion rationale.
- *Rules of engagement*: A process charter that captures facilitation rules, evidence standards and fallback procedures.
- *Appointment of facilitators*: A facilitator mandate note recording who is appointed, along with motivating reasons.

4.4.2 Communicative space

- *Orientations*: A briefing document with lay and technical summaries of the AI system under review, the dimensions of freedom and the SDG conversation starters.
- *Process and principles*: A roadmap with jointly agreed milestones and deliberation steps, aligned with mandate and limits defined in the coordinative space.
- *Definitions*: A working definition document with stakeholder-agreed definitions.
- *Judgements*: An argument log as a structured record of arguments, judgements and counter-judgements.

4.4.3 Decision space

- *Review*: A compliance note where facilitators confirm the fairness of process and interpreters confirm fair representation of inputs.
- *Synthesis*: A synthesis document that combines stakeholder judgements and highlights convergence and unresolved tensions.
- *Closure*: A formal decision record that documents how decision rules were used to reach final judgements, also including a record of dissenting views.
- *Reporting*: A public deliberation report which provides a publicly accessible summary of the process, argumentation and decisions.

### 4.5. Policymaker options

The intention of the framework is to guide policymakers in deciding how to respond to current, emerging or novel AI systems in their jurisdiction. But what are the options available to policymakers? To concretise available options, we have used Hood's NATO typology to map potential policymaker options. While not an exhaustive list, the intention is to provide a sense of the spread of options that could be considered after completion of the framework.

Table 2: Illustrative overview of policy instruments available to policymakers

| *Level of intervention* | **Nodality**<br>Knowledge and communication | **Authority**<br>Rules and regulations | **Treasure**<br>Economic incentives | **Organisation**<br>Government capacity |
|---|---|---|---|---|
| **Enabling** | Ethical charters, best-practice guides, public awareness campaigns | National AI strategies, voluntary impact assessment templates | Pilot grants, innovation prizes, seed funding | Temporary taskforces, advisory boards |
| **Steering** | Voluntary transparency registries, public scorecards on AI use | Technical standards for co- or self-regulation, certification schemes with soft enforcement | AI tax credits, subsidies for compliance tools | Semi-independent oversight bodies, AI observatories |
| **Compulsory** | Mandatory AI impact assessments (hybrid) | Binding legislation restricting harmful AI uses, mandatory AI impact assessments (hybrid) | Conditional grants tied to compliance, differentiated taxation | Permanent AI regulatory agencies, dedicated inspectorates |
| **Command** | State-run risk reporting dashboards, mandatory public disclosure portals | Constitutional or treaty-level commitments on AI rights and safeguards, AI procurement rules (hybrid) | AI procurement rules (hybrid), large-scale fiscal reallocation | State-owned AI infrastructure, in-house development of critical AI systems |

## 5. Conclusion

In this article it has been shown how a framework that uses freedom, rather than responsibility, and a deliberative, rather than technical, approach to policymaking can assist policymakers with deciding whether and how to respond to current, emerging and new AI systems.

In our reading, the framework makes a threefold contribution to the discourse on AI regulation. First, it shifts the focus of policymaking from responsibility and risk mitigation to societal benefit. Second, it moves from a technical to a deliberative policymaking rationality, which is in our view, better suited to conditions of uncertainty and contestation. Third, it provides a structured way to assess and debate the societal impact of AI systems, prioritising participation and transparency. Taken together, this contribution could strengthen policymakers' ability to respond appropriately to AI systems in their respective jurisdictions.